\documentclass[aps,prb,twocolumn,superscriptaddress,showpacs,showkeys,amsmath,amssymb]{revtex4}
\usepackage{amsfonts}
\usepackage{amssymb,amsmath}
\usepackage{graphicx}
\usepackage{dcolumn}
\usepackage{bm}
\usepackage{color}

\begin{document}

\title{Dispersion-driven ferromagnetism in a flat-band Hubbard system}

\author{Oleg Derzhko}
\affiliation{Max-Planck-Institut f\"{u}r Physik komplexer Systeme,
          N\"{o}thnitzer Stra\ss e 38, 01187 Dresden, Germany}
\affiliation{Institute for Condensed Matter Physics,
          National Academy of Sciences of Ukraine,
          1 Svientsitskii Street, L'viv-11, 79011, Ukraine}
\affiliation{Department for Theoretical Physics, 
          Ivan Franko National University of L'viv, 
          12 Drahomanov Street, L'viv-5, 79005, Ukraine}
\affiliation{Abdus Salam International Centre for Theoretical Physics,
          Strada Costiera 11, I-34151 Trieste, Italy}

\author{Johannes Richter}
\affiliation{Max-Planck-Institut f\"{u}r Physik komplexer Systeme,
          N\"{o}thnitzer Stra\ss e 38, 01187 Dresden, Germany}
\affiliation{Institut f\"{u}r theoretische Physik,
          Otto-von-Guericke-Universit\"{a}t Magdeburg,
          P.O. Box 4120, D-39016 Magdeburg, Germany}

\date{\today}






\begin{abstract}
We investigate a mechanism to establish ground-state ferromagnetism in flat-band Hubbard systems 
based on a kind of {\it order-from-disorder} effect driven by dispersion. 
As a paradigm we consider a frustrated diamond chain, 
where for ideal diamond geometry the lowest one-electron band is flat, 
but the ground state remains paramagnetic for arbitrary on-site repulsion $U$.
We focus on half filling of the flat band. 
By using numerical and analytical arguments we present the ground-state phase diagram for a  distorted diamond chain, 
i.e., the former flat band becomes dispersive. 
Driven by the interplay of  dispersion and interaction the ground state is ferromagnetic
if the interaction exceeds a critical value $U_c$.
\end{abstract}

\maketitle

Electronic as well as localized-spin systems with dispersionless (i.e., flat) one-particle bands 
offer a unique playground in view of realizing unconventional low-temperature phases, 
see, e.g., 
Refs.~\onlinecite{optical1,alter_1,topological, topological_tasaki, Heis_PRL,s_flach} 
and references therein.
An illustrious example is the fractional quantum Hall effect 
caused by interactions within the highly degenerate manifold of the dispersionless Landau levels.\cite{QHF} 
Very recently, 
remarkably in flat-band systems high-temperature fractional quantum Hall states without a magnetic field were found.\cite{topological}
Relevant perturbations leading to dispersion of the one-particle band or interactions between the particles 
may yield highly non-trivial correlation effects. 
An intriguing example is the ferromagnetic instability in flat-band Hubbard models  
when the interaction prevails against the kinetic energy.\cite{mielke,tasaki,mielke-tasaki,tasaki_flat_review,
1dtasaki_cerh3b2,prl2012}
This class  of interacting quantum systems belongs to the notoriously rare examples 
where rigorous results for many-body ground states are available. 
After the seminal papers by Mielke and Tasaki \cite{mielke,tasaki,mielke-tasaki,tasaki_flat_review}
a continuous theoretical activity in this field   
can be recorded.\cite{1dtasaki_cerh3b2,tasaki_jsp,kusakabe,kusakabe_robust,tasaki_nearly_flat,watanabe,tanaka-idogaki1, tanaka-idogaki2,tanaka_ueda,sekizawa,jmmm2004,prb2004,tanaka_tasaki,tasaki_epjb,topological_tasaki,
mielke1999,lu,mielke2012,batista,prb2007,prb2009,epjb2011,gulacsi,prl2012}
Additional interest in this subject stems from the experimental side.
Besides several promising materials realizing flat-band ferromagnetism (FM),\cite{tamura-quantum-dots-wires,nishino-3d-flatband, flat-band-experiment-polymers,lin-grapheneribbon,flat-band-graphene}
generated flat-band phases in optical lattices open a wide area for experimental research activities, 
see, e.g., Refs.~\onlinecite{bloch2005,bloch2008,chern2013}.

A crucial point to understand the mechanism leading to flat-band FM is the existence of localized eigenstates, 
i.e., the particles can be localized within small trapping cells on a lattice
see, e.g., Refs.~\onlinecite{Heis_PRL,mielke,tasaki,prb2007,prb2009,s_flach}.
Thus many-electron ground states can be constructed by filling the traps obeying the Pauli principle, 
and a geometrical representation of the localized many-particle states is at hand.
As a result relevant quantum degrees of freedom can be mapped on classical ones 
thus providing the powerful toolbox of classical statistical mechanics.\cite{prb2007,prb2009,tsunetsugu,unser_EPJB}
In addition, it is worth noticing that this mapping also opens a window 
to a new percolation problem, called Pauli-correlated percolation,
in which the first-order nature of an equilibrium percolation transition can be established.\cite{prl2012,shtengel}
These findings indicate that Hubbard flat-band systems are of broad conceptual relevance.

Based on the geometrical interpretation of the localized many-body states 
the saturated FM then corresponds to the full occupation  of all trapping cells 
say with up-electrons.\cite{prb2007} 
Importantly, 
it is necessary to have overlapping traps 
(i.e., neighboring cells must share at least one site).
Then even if all the cells are occupied with electrons having identical spin (symmetric spin state)
they can avoid the Hubbard on-site repulsion $U>0$ 
and this fully polarized state remains within the ground-state manifold.
However, in case that the trapping cells are isolated from each other there is no route to FM 
and the large set of paramagnetic states prevails against the ferromagnetic eigenstate.\cite{batista,prb2009,epjb2011,mielke2012}  

Introducing dispersion (violation of the flat-band geometry) 
typically modifies the balance of interaction and kinetic energy and tends to destabilize FM.
However, it was demonstrated by several studies 
that the ferromagnetic ground state  is robust 
if the flat band becomes (slightly) dispersive and the Hubbard repulsion $U$ is larger than a threshold $U_c>0$, 
where $U_c$ depends on the degree of the violation of the flat-band geometry.\cite{tasaki_jsp,kusakabe,kusakabe_robust}

In the present paper we will demonstrate 
that unexpectedly for a certain class of flat-band systems having isolated trapping cells 
just the dispersion, i.e., the kinetic energy, 
will open the route to ferromagnetic ground states.
This dispersion-driven FM resembles the celebrated {\it order-from-disorder} mechanism,\cite{villain,shender} 
i.e., due to distortions 
an ordered ground state is selected from the degenerate flat-band ground-state manifold where paramagnetic states dominate.

We consider the standard repulsive Hubbard model
\begin{equation}
\label{201}
H=\sum_{(ij),\sigma} t_{ij} \left(c^\dagger_{i,\sigma}c_{j,\sigma} + c^\dagger_{j,\sigma}c_{i,\sigma}\right)
+U\sum_i n_{i,\uparrow}n_{i,\downarrow}.
\end{equation}
Here the sum in the hopping term runs over all nearest neighbors on the lattice,
$\sigma=\uparrow,\downarrow$,
$t_{ij}>0$,
$c^\dagger_{i,\sigma}$ ($c_{i,\sigma}$) are the usual fermion creation (annihilation) operators,
the sum in the interaction term runs over all $N$ lattice sites,
and
$n_{i,\sigma}=c^\dagger_{i,\sigma}c_{i,\sigma}$.

\begin{figure}
\begin{center}
\includegraphics[clip=on,width=70mm,angle=0]{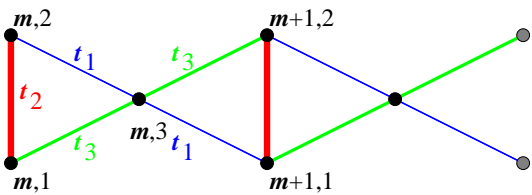}
\caption
{(Color online) The frustrated diamond chain corresponding to the Hamiltonian (\ref{201}).
The trapping cells for localized electrons (vertical dimers) are indicated by bold red lines ($t_2$ bonds).
The lattice sites are labeled by a pair of indeces, 
where the first number enumerates the unit cells
($m=1,\ldots,{\cal{N}}$, ${\cal{N}}=N/3$ is the number of unit cells)
and the second one enumerates the position of the site within the unit cell.}
\label{fig01}
\end{center}
\end{figure}

We focus on a generic example of the flat-band lattice with isolated trapping cells, 
namely  the frustrated diamond-chain lattice,\cite{prb2009} 
see Fig.~\ref{fig01}.
The flat-band case is realized for $t_1=t_3=t$.
Then electrons can be localized  within a characteristic trapping cell (vertical dimer) due to destructive quantum interference.
The one-particle energy of the flat-band states is $\varepsilon_1=-t_2$,
whereas the energies of the one-particle states from the two other dispersive bands are
$\varepsilon_{2,3}(\kappa)=t_2/2\mp\sqrt{t_2^2/4+4t^2(1+\cos\kappa)}$.
Here $\kappa=2\pi m/{\cal{N}}$, $m \in \mathbb{Z}$, $-{\cal{N}}/2\le m<{\cal{N}}/2$. 
The number of unit cells ${\cal{N}}= N/3$ equals the number of trapping cells.  
The two dispersive bands have the same widths 
$w_{2,3}=\sqrt{t_2^2/4+8t^2}-t_2/2\approx 2(t_3+t_1)^2/t_2$ if $t_2/t\gg 1$.
The flat-band states are the lowest-energy ones with a gap to excited states $\varepsilon_{2}(0)-\varepsilon_1>0$ 
if $t_2>2t$.

Owing to the localized nature of the one-electron flat-band states 
the many-electron states in the subspaces with $n=2,\ldots, {\cal{N}}$ electrons for $U>0$
can be constructed by filling the traps arbitrarily with up- or down-spin electrons 
leading to macroscopic ground state degeneracy of $2^{n}{\cal{C}}_{{\cal{N}}}^{n}$,
where the paramagnetic states are predominant.
Clearly  all these states are linearly independent.\cite{lin_indep}
Moreover, these localized many-electron states have the lowest energy in their corresponding $n$-electron subspaces, 
if the flat band is the lowest, i.e., for $t_2>2t$.

Now we focus on an electron filling corresponding to a half-filled flat band, i.e., $n={\cal{N}}$. 
Then all traps are occupied with precisely one electron with arbitrary spin orientation, 
and the intermediate sites out of the trapping cells are empty. 
The ground states exhibit perfect charge order, 
but the averaged magnetic moment per site at $T=0$ is zero for $N \to \infty$.\cite{prb2009}
Moreover, these charge-ordered ground states are protected by a charge gap.

A relevant violation of the ideal flat-band geometry is known from the fascinating frustrated magnetic compound azurite.\cite{azurite}
We adopt the azurite geometry and consider different values of $t_1$ and $t_3$ but fixing their average, 
i.e., $t_1\ne t_3$, $t_1+t_3=2t$, 
see Fig.~\ref{fig01}.
Moreover, we assume $t_2>2t$, see above.
For $t_1 \ne t_3$ the lowest-energy band acquires dispersion and its band width becomes $W_1\approx 2(t_3-t_1)^2/t_2$.
The ratio between $W_1$ and the band width $w_{2,3}$ of the upper dispersive bands
\begin{eqnarray}
\label{202}
\frac{W_1}{w_{2,3}}\approx \Omega^2,
\;\;\;
\Omega\equiv \left\vert\frac{t_3-t_1}{t_3+t_1}\right\vert
\end{eqnarray}
gives an appropriate dimensionless parameter $\Omega^2$ that characterizes the acquired dispersion of the former flat band
and which, in turn, can be used to measure the strength of deviation from the ideal flat-band geometry. 

\begin{figure}
\begin{center}
\includegraphics[clip=on,width=85mm,angle=0]{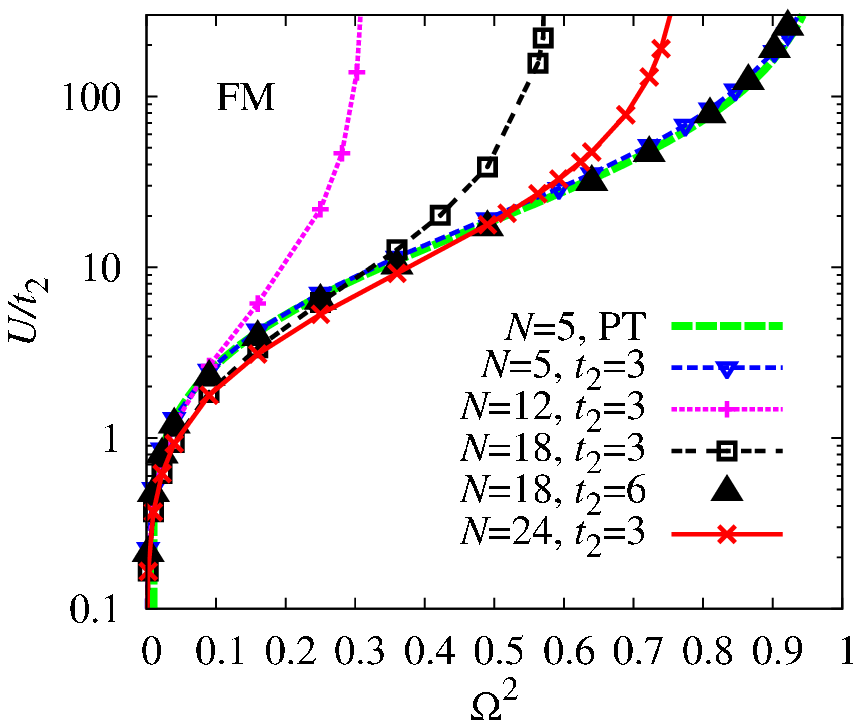}\\
\vspace{5mm}
\includegraphics[clip=on,width=85mm,angle=0]{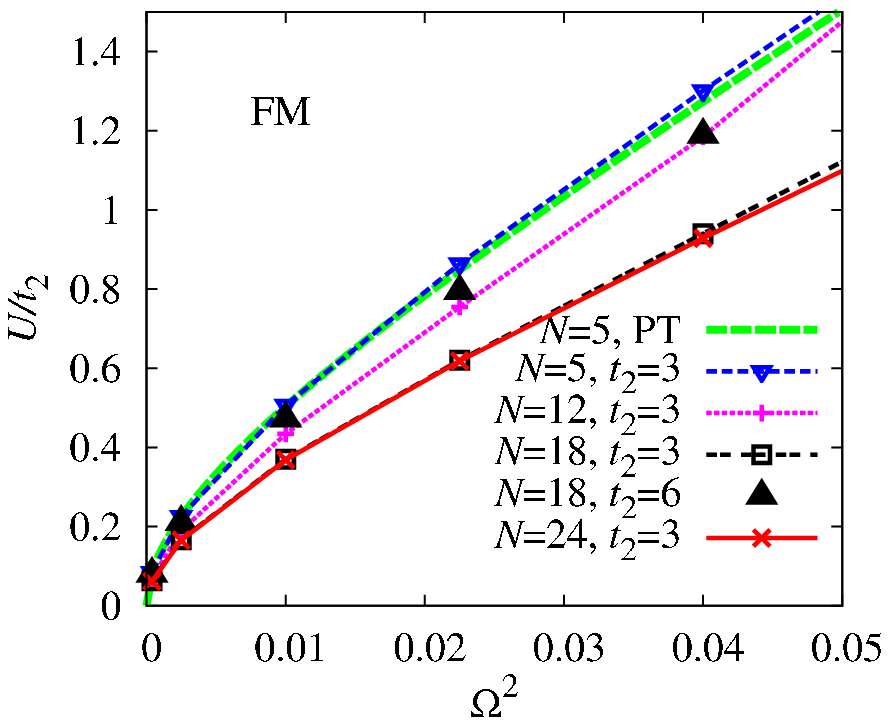}
\caption
{(Color online) 
Phase diagram of the distorted Hubbard diamond chain with electron density  $n/N=1/3$ (half-filled lowest band).
Ferromagnetism (denoted by ``FM'') appears for on-site repulsions $U$ above a critical value $U_c$. 
$U_c$ is shown as a function of the dimensionless band-width parameter $\Omega^2$, see Eq.~(\ref{202}).
The various critical lines $U_c(\Omega^2)$ are obtained 
by fourth-order perturbation theory (symbol ``PT''), see Eq.~(\ref{303}), 
and by exact diagonalization
for finite systems of $N=5$ (open boundary conditions) and $N=12,18,24$ (periodic boundary conditions) for hopping parameter sets with $t_3+t_1=2$ and  $t_2=3,6$.}
\label{fig02}
\end{center}
\end{figure}

Our main result obtained by exact diagonalization (ED) and fourth-order perturbation theory (PT) 
is compactly illustrated by the phase diagram in Fig.~\ref{fig02}.
For not too large deviations from ideal geometry controlled by $\Omega^2<1$ and for sufficiently large $U/t_2>U_c/t_2$, 
the ground state of the Hubbard diamond chain with half-filled lowest band is ferromagnetic 
(the region denoted as ``FM'').

Let us now illustrate briefly our calculations leading to Fig.~\ref{fig02}.
The Lanczos ED to calculate the ground states was performed for chains of $N=12,18,24$ sites,
i.e., of ${\cal{N}}=4,6,8$ unit cells, 
with periodic boundary conditions imposed.
We set $t_3+t_1=2$ and consider various values of $t_2$. 
In  Fig.~\ref{fig02} we  present data for $t_2=3$ and $6$.
The parameter characterizing the dispersion was varied from $t_3-t_1=0$ ($\Omega^2=0$) to $t_3-t_1= 2$ ($\Omega^2=1$).
For quite (but not arbitrarily) large deviations from ideal geometry $\Omega^2$ and sufficiently large $U>U_c$
there is a unique ferromagnetic ground state, 
i.e., the ground state is a ferromagnetic SU(2) multiplet which has the degeneracy $n+1={\cal{N}}+1$. 
In the limit  $U \to \infty$ there is a maximal dispersion $\Omega_c^2$
above  which ground-state FM does not exist.
As expected there is a finite-size dependence of $U_c(\Omega^2)$, 
but the general shape of the curve is the same for all $N$.  
Interestingly the region of FM increases  with growing system size $N$. 
Thus, for $N=12, 18, 24$, and $30$ sites with $t_2=3$  
we found in the limit $U\to \infty$ the values $\Omega_c^2\approx 0.354, 0.591, 0.778$, and $0.884$, respectively.
That gives clear evidence that the dispersion-driven FM exists for $N \to \infty$.
The influence of the vertical hopping integral $t_2$ on the phase boundary $U_c(\Omega^2)$ 
is visible from the corresponding curves for $N=18$. 
For small band width $\Omega^2$ the region of FM slightly shrinks with increasing of $t_2$, 
whereas $U_c$ grows with increasing $t_2$ for $\Omega^2  \gtrsim 0.25$.

Let us finally mention, 
that the dispersion does not change substantially the charge order present for the ideal flat-band geometry.
For example, for $N=18$, $U=10$, $t_2=3$, $\Omega^2=0.25$ 
the occupation of the intermediate sites amounts less than 10\% of the occupation of the sites of the trapping cells.

In the next step we complement our numerical data by fourth-order PT. 
We will illustrate only some main features of this approach.
More details can be found in the Appendix.
In fourth order it is sufficient to consider a $5$-site cluster that contains two trapping cells 
(the sites $m,1$, $m,2$, $m,3$, $m+1,1$, and $m+1,2$ connected by six bonds in Fig.~\ref{fig01}) 
in the subspace of $n=2$ electrons.
We apply standard PT.\cite{klein,fulde,essler,harris} 
The unperturbed Hamiltonian contains 
just the $t_2$- and $U$-terms (recall that $t_2>2t=t_3+t_1$ is the dominating hopping integral). 
The perturbation contains the $t_3$- and $t_1$-terms.
The unperturbed ground states in the two-electron subspace with accounting SU(2) symmetry have the form
$\vert t,1\rangle=l_{m,\uparrow}^\dagger l_{m+1,\uparrow}^\dagger \vert 0\rangle$,
$\vert t,0\rangle
=
(1/\sqrt{2})
(l_{m,\uparrow}^\dagger l_{m+1,\downarrow}^\dagger
     +l_{m,\downarrow}^\dagger l_{m+1,\uparrow}^\dagger)\vert 0\rangle$,
$\vert t,-1\rangle=l_{m,\downarrow}^\dagger l_{m+1,\downarrow}^\dagger \vert 0\rangle$
(the components of the triplet state),
and
$\vert s\rangle
=
(1/\sqrt{2})(l_{m,\uparrow}^\dagger l_{m+1,\downarrow}^\dagger
     -l_{m,\downarrow}^\dagger l_{m+1,\uparrow}^\dagger)\vert 0\rangle$
(the singlet state).
Here $l^{\dagger}_{m,\sigma}=(c^{\dagger}_{m,1,\sigma}-c^{\dagger}_{m,2,\sigma})/\sqrt{2}$
and $\vert 0\rangle$ denotes the vacuum state.

Calculating the energy up to the fourth order we find for the triplet states
\begin{eqnarray}
\label{301}
 E_{t}=
-2t_2-\frac{\left(t_3-t_1\right)^2}{t_2}
\nonumber\\
-\frac{\left(t_3+t_1\right)^2\left(t_3-t_1\right)^2}{2t_2^3}+\frac{\left(t_3-t_1\right)^4}{t_2^3}
+\ldots ,
\end{eqnarray}
i.e., $E_t$ is independent of $U$.
The  energy of the singlet state $E_s$ depends on $U$.   
For the forth-order analytical expression for $E_s(U)$ see the Appendix.
In the limit $U\to\infty$ it becomes
\begin{eqnarray}
\label{302}
 E_{s}(\infty)=
-2t_2-\frac{\left(t_3-t_1\right)^2}{t_2}
\nonumber\\
-\frac{\left(t_3+t_1\right)^2\left(t_3-t_1\right)^2}{4t_2^3}+\frac{3\left(t_3-t_1\right)^4}{4t_2^3}
+\ldots
\end{eqnarray}
and $E_{s}(\infty)>E_t$ due to the fourth-order term.
Thus, the ground state in the limit $U\to\infty$ is ferromagnetic.
In the small-$U$ limit,
the dominating term in $E_s(U)$ is $-2(t_3-t_1)^4/(Ut_2^2)$
which obviously leads to the opposite inequality $E_{s}(U)<E_t$, 
i.e., the ground state is nonmagnetic.
To determine the critical value of $U_c$, above which the ground state is
ferromagnetic we have to solve the equation $E_{t}=E_{s}(U_c)$
which gives
\begin{eqnarray}
\label{303}
\frac{U_c}{t_2}=\frac{\sqrt{16+65\Omega^2}+9\Omega}{1-\Omega^2}\Omega.
\end{eqnarray}
Obviously, for small deviations from the flat-band case, i.e., $\Omega^2 \ll 1$, we get $U_c/t_2 \approx 4\Omega$.
The graphical representation of Eq.~(\ref{303}) is shown by the thick long-dashed green line in Fig.~\ref{fig02}.
The effective Hamiltonian describing the low-energy degrees of freedom of the 5-site two-electron Hubbard problem
is the 2-site spin-1/2 Heisenberg model
\begin{eqnarray}
\label{304}
H_{\rm{eff}}={\sf{J}}(U){\bf{T}}_m\cdot{\bf{T}}_{m+1} + {\sf{C}}(U),
\end{eqnarray}
where
${\sf{J}}(U)=E_t-E_s(U)$,
${\sf{C}}(U)=[3E_t+E_s(U)]/4$.
The pseudospin operators are given by
$T_m^+=l^{\dagger}_{m,\uparrow} l_{m,\downarrow}$,
$T_m^-=l^{\dagger}_{m,\downarrow} l_{m,\uparrow}$,
and
$T_m^z=(l^{\dagger}_{m,\uparrow} l_{m,\uparrow}-l^{\dagger}_{m,\downarrow} l_{m,\downarrow})/2$.
The exchange constant ${\sf{J}}(U)$ in Eq.~(\ref{304}) is positive in the small-$U$ limit,
changes its sign at $U_c$ given by Eq.~(\ref{303}),
and approaches $-t_3t_1(t_3-t_1)^2/t_2^3<0$ as $U\to\infty$.
According to the PT the phase diagram is universal if we use $\Omega^2$ and $U/t_2$ for the axis,
i.e., the data for $U_c$ for various $t_1$, $t_2$, $t_3$ should collapse to one universal curve (\ref{303}).
Another consequence of Eq.~(\ref{303}) is that $U_c\to\infty$ if $\Omega^2\to 1$, 
i.e., Eq.~(\ref{303}) yields $\Omega^2_c=1$.
As a direct check of the fourth-order PT results for $E_t$ and $E_s(U)$ we have performed ED for the $5$-site cluster, 
see the dashed blue line with empty down-triangles in Fig.~\ref{fig02}.
The agreement is excellent.

In summary, the PT confirms our ED results for the existence of FM. 
The quantitative agreement between ED and PT is the better the smaller $\Omega^2$ and the larger $t_2$. 
The appearance of ground-state FM for deviations from ideal geometry is a result of fourth-order processes.
Higher-order PT naturally would enlarge the region of quantitative agreement with ED data.

\begin{figure}
\begin{center}
\includegraphics[clip=on,width=85mm,angle=0]{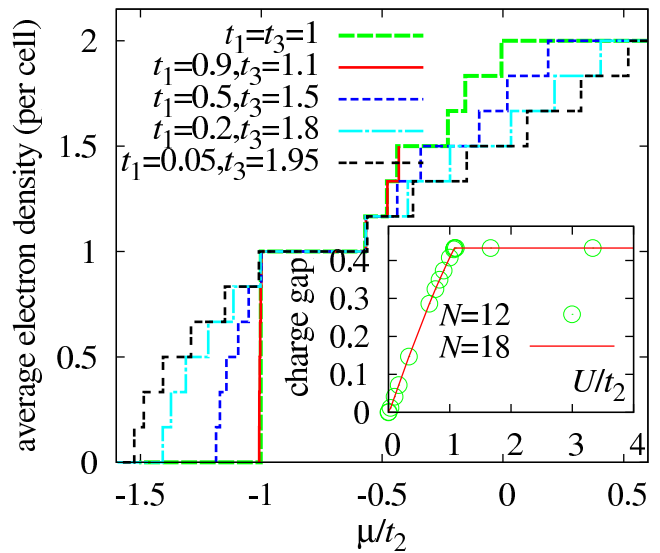}
\caption
{(Color online) 
Averaged electron density (per cell) $\overline{n}/{\cal{N}}$ versus chemical potential $\mu/t_2$
for the frustrated diamond chain of $N=18$ sites with $t_2=3$, 
$t_1=t_3=1$, 
$t_1=0.9$, $t_3=1.1$, 
$t_1=0.5$, $t_3=1.5$,
$t_1=0.2$, $t_3=1.8$,
and
$t_1=0.05$, $t_3=1.95$,   
and $U\to\infty$.
Inset: Charge gap (plateau width) $\Delta\mu/t_2$ at $\overline{n}/{\cal{N}}=1$ versus $U/t_2$ 
for the distorted frustrated diamond chains of $N=12$ (circles) and $N=18$ (line) sites 
with $t_2=3$, $t_1=0.9$, $t_3=1.1$.}
\label{fig03}
\end{center}
\end{figure}

For experimental research 
the existence of a charge gap that would protect the considered state against charge fluctuations is of relevance. 
As already mentioned above such a charge gap exists for the ideal geometry.\cite{prb2009}
To calculate the charge gap for the distorted geometry 
we consider the grand-canonical setup adding to the Hamiltonian (\ref{201}) the term with a chemical potential, 
$-\mu\sum_i (n_{i,\uparrow}+n_{i,\downarrow})$.
In Fig.~\ref{fig03} we show the ground-state behavior of the averaged electron density (per cell) $\overline{n}/{\cal{N}}$ 
as a function  of the chemical potential $\mu$.  
There is a wide plateau appearing for $U \gtrsim 1$ at $\overline{n}/{\cal{N}}=1$ that is almost independent of the distortion. 
For  $U \lesssim 1$ the plateau is less pronounced and it disappears for $U=0$.
The width of the plateau $\Delta\mu$ corresponds to the size of the charge gap. Fig.~\ref{fig03} gives evidence 
that this charge gap is robust against the deviation from ideal flat-band geometry, 
and, thus, the protection works also in this case.
Moreover, the finite-size dependence is small (see also Refs.~\onlinecite{prb2007,prb2009}).

To summarize,
we have considered a dispersion-driven emergence 
of the ground-state FM in the Hubbard model on a frustrated diamond chain
using ED and PT approaches.
The ferromagnetic ground state is observed for half filling of the lowest one-particle band. 
It is protected by a charge gap against charge fluctuations and exhibits also charge ordering.
The discussed scenario of dispersion-driven FM
is not restricted to specific geometry of the frustrated diamond chain,
rather it is quite general for lattices with isolated cells including two-dimensional ones.
Indeed ED calculations confirm this expectations for the lattices considered in Ref.~\onlinecite{prb2009}.

Our focus on the diamond-chain geometry is motivated by the fact that this geometry is often found in natural compounds,
see, e.g., Refs.~\onlinecite{azurite,diamondchains_ssrealizations,honeck}.
Its simplicity is advantageous in realizing this geometry, 
e.g., in optical lattices or quantum dot arrays.
Since one may expect that it is hard to realize the perfect flat-band geometry in experiments 
the discussion of distortions is of crucial relevance.

There is an ongoing experimental search for flat-band ferromagnets.
The possibility of flat-band FM in quantum dot arrays and in quantum atomic wires formed on solid surfaces 
was discussed in Ref.~\onlinecite{tamura-quantum-dots-wires}.
Other examples for the flat-band ferromagnets come from polymers\cite{flat-band-experiment-polymers}
where a search for purely organic ferromagnets is known as a challenging target.
Another experimental realization 
of the flat-band FM has been discussed recently in Ref.~\onlinecite{flat-band-graphene}.
Specific organic molecules 
[tetracyano-$p$-quinodimethane (TCNQ) molecules] 
deposited on graphene epitaxially grown on Ru(0001) 
acquire charge from the substrate and develop a magnetic moment which survives when the molecules form a monolayer.
The authors of Ref.~\onlinecite{flat-band-graphene} claim that the intermolecular bands are almost flat and half-filled
and that the TCNQ/graphene/Ru(0001) system might be a realization of the flat-band Hubbard ferromagnets.

Our theoretical study provides a further route in search of flat-band ferromagnets:
For this purpose one can use not only Mielke's or Tasaki's flat-band systems.\cite{mielke,tasaki} 
Also flat-band systems which do not show ground-state FM for ideal geometry are promising candidates, 
since one may expect that deviations from the perfect flat-band geometry 
are always present in experimental realizations of such systems.

\section*{Acknowledgments}

We are grateful to M.~Maksymenko for discussions.
The numerical calculations were performed using J.~Schulenburg's {\it spinpack}.\cite{spinpack}
The present study was supported by the DFG (project RI615/21-1).
J.~R. and O.~D. would like to acknowledge the hospitality of the MPIPKS, Dresden in October-December of 2013.
O.~D. would like to thank the Abdus Salam International Centre for Theoretical Physics (Trieste, Italy) 
for partial support of these studies through the Senior Associate award.

\onecolumngrid

\section*{Appendix: Perturbation theory for the 5-site two-electron problem}
\renewcommand{\theequation}{A\arabic{equation}}
\setcounter{equation}{0}

In this Appendix
we present some details of the perturbation-theory calculations for the 5-site two-electron problem
which are used in the main text.

We split the Hamiltonian of the model $H$ given in Eq.~(\ref{201}) for the 5-site cluster into the main part
\begin{eqnarray}
\label{a01}
{\sf{H}}_0=\sum_{\sigma=\uparrow,\downarrow}
\left[t_2\left(c^\dagger_{m,1,\sigma}c_{m,2,\sigma}+c^\dagger_{m,2,\sigma}c_{m,1,\sigma}\right)
+t_2\left(c^\dagger_{m+1,1,\sigma}c_{m+1,2,\sigma}+c^\dagger_{m+1,2,\sigma}c_{m+1,1,\sigma}\right)\right]
\nonumber\\
+U\left(n_{m,1,\uparrow}n_{m,1,\downarrow}+n_{m,2,\uparrow}n_{m,2,\downarrow}+n_{m,3,\uparrow}n_{m,3,\downarrow}
+n_{m+1,1,\uparrow}n_{m+1,1,\downarrow}+n_{m+1,2,\uparrow}n_{m+1,2,\downarrow}\right)
\end{eqnarray}
and the perturbation
\begin{eqnarray}
\label{a02}
{\sf{V}}=\sum_{\sigma=\uparrow,\downarrow}
\left[
t_3\left(c^\dagger_{m,1,\sigma}c_{m,3,\sigma}+c^\dagger_{m,3,\sigma}c_{m,1,\sigma}\right)
+t_1\left(c^\dagger_{m,2,\sigma}c_{m,3,\sigma}+c^\dagger_{m,3,\sigma}c_{m,2,\sigma}\right)
\right.
\nonumber\\
\left.
+t_1\left(c^\dagger_{m,3,\sigma}c_{m+1,1,\sigma}+c^\dagger_{m+1,1,\sigma}c_{m,3,\sigma}\right)
+t_3\left(c^\dagger_{m,3,\sigma}c_{m+1,2,\sigma}+c^\dagger_{m+1,2,\sigma}c_{m,3,\sigma}\right)
\right].
\end{eqnarray}
Using as a complete set of the one-electron states the states
$l^\dagger_{m,\sigma}\vert 0\rangle$,
$l^\dagger_{m+1,\sigma}\vert 0\rangle$,
$c^\dagger_{m,3,\sigma}\vert 0\rangle$,
$d^\dagger_{m,\sigma}\vert 0\rangle$,
$d^\dagger_{m+1,\sigma}\vert 0\rangle$
($\sigma=\uparrow,\downarrow$)
with 
$l^\dagger_{m,\sigma}=(c^\dagger_{m,1,\sigma}-c^\dagger_{m,2,\sigma})/\sqrt{2}$,
$d^\dagger_{m,\sigma}=(c^\dagger_{m,1,\sigma}+c^\dagger_{m,2,\sigma})/\sqrt{2}$,
we find all 45 eigenstates $\vert\alpha\rangle$ and their energies ${\sf{E}}_{\alpha}$
of the unperturbed Hamiltonian ${\sf{H}}_0$ (\ref{a01}) in the two-electron subspace.
The ground state of the unperturbed Hamiltonian ${\sf{H}}_0$ in the two-electron subspace $\vert{\rm{GS}}\rangle$
is four-fold degenerate,
i.e., consists of the 3 triplet states, 
$\vert t,1\rangle=l^{\dagger}_{m,\uparrow}l^{\dagger}_{m+1,\uparrow}\vert 0\rangle$, 
$\vert t,0\rangle=(1/\sqrt{2})(l^{\dagger}_{m,\uparrow}l^{\dagger}_{m+1,\downarrow}+l^{\dagger}_{m,\downarrow}l^{\dagger}_{m+1,\uparrow})\vert 0\rangle$,  
$\vert t,-1\rangle=l^{\dagger}_{m,\downarrow}l^{\dagger}_{m+1,\downarrow}\vert 0\rangle$,
and the singlet state,
$\vert s\rangle=(1/\sqrt{2})(l^{\dagger}_{m,\uparrow}l^{\dagger}_{m+1,\downarrow}-l^{\dagger}_{m,\downarrow}l^{\dagger}_{m+1,\uparrow})\vert 0\rangle$,  
with the ground-state energy ${\sf{E}}_{{\rm{GS}}}=E_t^{(0)}=E_s^{(0)}=-2t_2$.

The lowest-order perturbation-theory corrections to the ground-state energy ${\sf{E}}_{{\rm{GS}}}$ are as follows:
\begin{eqnarray}
\label{a03}
E^{(2)}_{{\rm{GS}}}=\sideset{}{'}\sum_{\alpha}
\frac{\langle{\rm{GS}}\vert{\sf{V}}\vert\alpha\rangle\langle\alpha\vert{\sf{V}}\vert{\rm{GS}}\rangle}
{{\sf{E}}_{{\rm{GS}}}-{\sf{E}}_{\alpha}},
\nonumber\\
E^{(3)}_{{\rm{GS}}}=\sideset{}{'}\sum_{\alpha} \sideset{}{'}\sum_{\beta}
\frac{\langle{\rm{GS}}\vert{\sf{V}}\vert\alpha\rangle\langle\alpha\vert{\sf{V}}\vert\beta\rangle
\langle\beta\vert{\sf{V}}\vert{\rm{GS}}\rangle}
{\left({\sf{E}}_{{\rm{GS}}}-{\sf{E}}_{\alpha}\right)\left({\sf{E}}_{{\rm{GS}}}-{\sf{E}}_{\beta}\right)},
\nonumber\\
E^{(4)}_{{\rm{GS}}}=\sideset{}{'}\sum_{\alpha} \sideset{}{'}\sum_{\beta} \sideset{}{'}\sum_{\gamma}
\frac{\langle{\rm{GS}}\vert{\sf{V}}\vert\alpha\rangle\langle\alpha\vert{\sf{V}}\vert\beta\rangle
\langle\beta\vert{\sf{V}}\vert\gamma\rangle\langle\gamma\vert{\sf{V}}\vert{\rm{GS}}\rangle}
{\left({\sf{E}}_{{\rm{GS}}}-{\sf{E}}_{\alpha}\right)\left({\sf{E}}_{{\rm{GS}}}-{\sf{E}}_{\beta}\right)
\left({\sf{E}}_{{\rm{GS}}}-{\sf{E}}_{\gamma}\right)}
-\sideset{}{'}\sum_{\alpha} \sideset{}{'}\sum_{\beta}
\frac{\langle{\rm{GS}}\vert{\sf{V}}\vert\alpha\rangle\langle\alpha\vert{\sf{V}}\vert{\rm{GS}}\rangle
\langle{\rm{GS}}\vert{\sf{V}}\vert\beta\rangle\langle\beta\vert{\sf{V}}\vert{\rm{GS}}\rangle}
{\left({\sf{E}}_{{\rm{GS}}}-{\sf{E}}_{\alpha}\right)^2\left({\sf{E}}_{{\rm{GS}}}-{\sf{E}}_{\beta}\right)}.
\end{eqnarray}
Here the superscript prime means 
that the sum extends over all states of the unperturbed Hamiltonian ${\sf{H}}_0$ except the ground states.
After straightforward calculations on the base of Eqs.~(\ref{a03}) and (\ref{a02})
we find the following nonzero corrections to the ground-state energy:
\begin{eqnarray}
\label{a04}
E^{(2)}_{t}=E^{(2)}_{s}=-\frac{\left(t_3-t_1\right)^2}{t_2}
\end{eqnarray}
and
\begin{eqnarray}
\label{a05a}
E^{(4)}_{t}=-\frac{\left(t_3+t_1\right)^2\left(t_3-t_1\right)^2}{2t_2^3}+\frac{\left(t_3-t_1\right)^4}{t_2^3},
\end{eqnarray}
\begin{eqnarray}
\label{a05b}
E^{(4)}_{s}(U)=-\frac{\left(t_3+t_1\right)^2\left(t_3-t_1\right)^2}{4t_2^3}
+\frac{\left(t_3-t_1\right)^4}{t_2^3}
\nonumber\\
-\frac{a^2\left(t_3-t_1\right)^4}{\left(2t_2+\frac{U}{2}-\sqrt{\frac{U^2}{4}+4t_2^2}\right)t_2^2}
-\frac{\left(t_3+t_1\right)^2\left(t_3-t_1\right)^2}{2\left(2t_2+U\right)t_2^2}
-\frac{2\left(t_3-t_1\right)^4}{\left(2t_2+U\right)t_2^2}
-\frac{b^2\left(t_3-t_1\right)^4}{\left(2t_2+\frac{U}{2}+\sqrt{\frac{U^2}{4}+4t_2^2}\right)t_2^2},
\nonumber\\
a^2=\frac{1}{C_-^2}\left(\frac{U}{4t_2}-\sqrt{\frac{U^2}{16t_2^2}+1}-1\right)^2,
\;\;\;
b^2=\frac{1}{C_+^2}\left(\frac{U}{4t_2}+\sqrt{\frac{U^2}{16t_2^2}+1}-1\right)^2,
\nonumber\\
C^2_{\mp}
=
\left(\frac{U}{4t_2}\mp\sqrt{\frac{U^2}{16t_2^2}+1}-1\right)^2
+
\left(\frac{U}{4t_2}\mp\sqrt{\frac{U^2}{16t_2^2}+1}+1\right)^2,
\end{eqnarray}
i.e., the ground-state degeneracy is lifted in the fourth order of the
perturbation theory.

The triplet-state energy does not depend on $U$;
the obtained results (\ref{a04}), (\ref{a05a}) are given in Eq.~(\ref{301}).
In the limit $U\to\infty$, 
we have $U/2-\sqrt{U^2/4+4t_2^2}\to 0$, $a^2\to 1/2$
and the singlet-state energy $E_s(\infty)$ is given by the formula in Eq.~(\ref{302}).
In the small-$U$ limit, 
we have $\sqrt{U^2/4+4t_2^2}\to 2t_2$, $a^2\to1$, $b^2\to 0$ 
and the dominating term in $E_s(U)$ (\ref{a05b}) is $-2(t_3-t_1)^4/(Ut_2^2)$.

Using Eqs.~(\ref{a05a}) and (\ref{a05b}), the equation for $U_c$, $E_t=E_s(U_c)$, can be written as follows:
\begin{eqnarray}
\label{a06}
\left(\frac{t_3-t_1}{t_3+t_1}\right)^2=\frac{\frac{U_c^2}{t_2^2}}{\frac{U_c^2}{t_2^2}+18\frac{U_c}{t_2}+16}.
\end{eqnarray}
Solving Eq.~(\ref{a06}) with respect to $U_c/t_2$ we get Eq.~(\ref{303}).


\begin{thebibliography}{99}

\bibitem{topological}
D.~N.~Sheng, Z.-C.~Gu, K.~Sun, and L.~Sheng,
Nat. Commun. {\bf 2}, 389 (2011);
E.~Tang, J.-W.~Mei, and X.-G.~Wen,
Phys. Rev. Lett. {\bf 106}, 236802 (2011);
K.~Sun, Z.~Gu, H.~Katsura, and S.~Das Sarma,
Phys. Rev. Lett. {\bf 106}, 236803 (2011);
T.~Neupert, L.~Santos, C.~Chamon, and C.~Mudry,
Phys. Rev. Lett. {\bf 106}, 236804 (2011);
E.~Bergholtz and Zhao Liu,
Int. J. Mod. Phys. B {\bf 27}, 1330017 (2013).

\bibitem{topological_tasaki}
H.~Katsura, I.~Maruyama, A.~Tanaka, and H.~Tasaki,
Europhys. Lett. {\bf 91}, 57007 (2010).

\bibitem{optical1}
C.~Wu, D.~Bergman, L.~Balents, and S.~Das Sarma,
Phys. Rev. Lett. {\bf 99}, 070401 (2007);
Y.-F.~Wang, Z.-C.~Gu, C.-D.~Gong, and D.~N.~Sheng,
Phys. Rev. Lett. {\bf 107}, 146803 (2011).

\bibitem{alter_1}
Z.~Gul\'{a}csi, A.~Kampf, and D.~Vollhardt,
Phys. Rev. Lett. {\bf 105}, 266403 (2010).

\bibitem{Heis_PRL}
J.~Schulenburg, A.~Honecker, J.~Schnack, J.~Richter, and H.-J.~Schmidt,
Phys. Rev. Lett. {\bf 88}, 167207 (2002);
J.~Richter, J.~Schulenburg, A.~Honecker, J.~Schnack, and H.-J.~Schmidt,
J. Phys.: Condens. Matter {\bf 16}, S779 (2004);
J.~Richter, O.~Derzhko, and J.~Schulenburg,
Phys. Rev. Lett. {\bf 93}, 107206 (2004).

\bibitem{s_flach}
D.~Leykam, S.~Flach, O.~Bahat-Treidel, and A.~S.~Desyatnikov,
Phys. Rev. B {\bf 88}, 224203 (2013);
S.~Flach, D.~Leykam, J.~D.~Bodyfelt, P.~Matthies, and A.~S.~Desyatnikov,
Europhys. Lett. {\bf 105}, 30001 (2014).

\bibitem{QHF} 
D.~C.~Tsui, H.~L.~St\"{o}rmer, and A.~C.~Gossard, 
Phys. Rev. Lett. {\bf 48}, 1559 (1982);
H.~L.~St\"{o}rmer, 
Rev. Mod. Phys. {\bf 71}, 875 (1999).

\bibitem{mielke}
A.~Mielke, 
J. Phys. A {\bf 24}, L73 (1991);
{\bf 24}, 3311 (1991);
{\bf 25}, 4335 (1992);
Phys. Lett. A {\bf 174}, 443 (1993).

\bibitem{tasaki}
H.~Tasaki,
Phys. Rev. Lett. {\bf 69}, 1608 (1992).

\bibitem{mielke-tasaki}
A.~Mielke and H.~Tasaki, 
Commun. Math. Phys. {\bf 158}, 341 (1993).

\bibitem{tasaki_flat_review}
H.~Tasaki, 
Prog. Theor. Phys. {\bf 99}, 489 (1998).

\bibitem{1dtasaki_cerh3b2}
T.~Okubo, M.~Yamada, A.~Thamizhavel, S.~Kirita, Y.~Inada, R.~Settai, 
H.~Harima, K.~Takegahara, A.~Galatanu, E.~Yamamoto, and Y.~Onuki,
J. Phys.: Condens. Matter {\bf 15},  L721 (2003); 
Z.~Gul\'{a}csi, A.~Kampf, and D.~Vollhardt, 
Phys. Rev. Lett. {\bf 99}, 026404 (2007).

\bibitem{prl2012} 
M.~Maksymenko, A.~Honecker,  R.~Moessner, J.~Richter,  and O.~Derzhko,
Phys. Rev. Lett. {\bf 109}, 096404 (2012).

\bibitem{tasaki_jsp}
H.~Tasaki, 
Phys. Rev. Lett. {\bf 73}, 1158 (1994);
{\bf 75}, 4678 (1995);
Commun. Math. Phys. {\bf 242}, 445 (2003).

\bibitem{kusakabe}
K.~Kusakabe and H.~Aoki,
Phys. Rev. Lett. {\bf 72}, 144 (1994);
H.~Aoki,
Int. J. Mod. Phys. B {\bf 17}, 4953 (2003).

\bibitem{kusakabe_robust}
K.~Kusakabe and H.~Aoki,
Physica B {\bf 194-196}, 215 (1994).

\bibitem{tasaki_nearly_flat}
H.~Tasaki, J. Stat. Phys. {\bf 84}, 535 (1996).

\bibitem{watanabe}
Y.~Watanabe and S.~Miyashita, 
J. Phys. Soc. Jpn. {\bf 66}, 2123 (1997); 
{\bf 66}, 3981 (1997); 
{\bf 68}, 3086 (1999); 
R.~Arita and H.~Aoki, 
Phys. Rev. B {\bf 61}, 12261 (2000).

\bibitem{tanaka-idogaki1}
A.~Tanaka and T.~Idogaki,
J. Phys. Soc. Jpn. {\bf 67}, 401 (1998). 

\bibitem{mielke1999}
A.~Mielke, 
Phys. Rev. Lett. {\bf 82}, 4312 (1999); 
J. Phys. A {\bf 32}, 8411 (1999).

\bibitem{tanaka-idogaki2}
A.~Tanaka and T.~Idogaki,
Physica A {\bf 297}, 441 (2001).

\bibitem{tanaka_ueda}
A.~Tanaka and H.~Ueda, 
Phys. Rev. Lett. {\bf 90}, 067204 (2003).

\bibitem{sekizawa}
T.~Sekizawa, 
J. Phys. A {\bf 36}, 10451 (2003).

\bibitem{batista}
C.~D.~Batista and B.~S.~Shastry, 
Phys. Rev. Lett. {\bf 91}, 116401 (2003).

\bibitem{jmmm2004}
H.~Ueda, A.~Tanaka, and T.~Idogaki,
J. Magn. Magn. Mater. {\bf 272-276}, 950 (2004).

\bibitem{prb2004}
H.~Ueda and T.~Idogaki,
Phys. Rev. B {\bf 69}, 104424 (2004).

\bibitem{tanaka_tasaki}
A.~Tanaka and H.~Tasaki, 
Phys. Rev. Lett. {\bf 98}, 116402 (2007).

\bibitem{gulacsi}
Z.~Gul\'{a}csi, A.~Kampf, and D.~Vollhardt, 
Prog. Theor. Phys. Suppl. 176, 1 (2008);
R.~Trencs\'{e}nyi, E.~Kov\'{a}cs, and Z.~Gul\'{a}csi, 
Philos. Mag. {\bf 89}, 1953 (2009);
R.~Trencs\'{e}nyi and Z.~Gul\'{a}csi,
Eur. Phys. J. B {\bf 75}, 511 (2010);
Z.~Gul\'{a}csi, 
Int. J. Mod. Phys. B {\bf 27}, 1330009 (2013).

\bibitem{prb2007}
O.~Derzhko, A.~Honecker, and J.~Richter,
Phys. Rev. B {\bf 76}, 220402 (2007);
O.~Derzhko, J.~Richter, A.~Honecker, M.~Maksymenko, and R.~Moessner,
Phys. Rev. B {\bf 81}, 014421 (2010).

\bibitem{tasaki_epjb}
H.~Tasaki,
Eur. Phys. J. B {\bf 64}, 365 (2008).

\bibitem{prb2009}
O.~Derzhko, A.~Honecker, and J.~Richter,
Phys. Rev. B {\bf 79}, 054403 (2009).

\bibitem{lu}
L.~Lu,
J. Phys. A {\bf 42}, 265002 (2009).

\bibitem{epjb2011}
M.~Maksymenko, O.~Derzhko, and J.~Richter,
Eur. Phys. J. B {\bf 84}, 397 (2011).

\bibitem{mielke2012}
A.~Mielke,
Eur. Phys. J. B {\bf 85}, 184 (2012).

\bibitem{tamura-quantum-dots-wires}
H.~Tamura, K.~Shiraishi, T.~Kimura, and H.~Takayanagi, 
Phys. Rev. B {\bf 65}, 085324 (2002);
T.~Kimura, H.~Tamura, K.~Shiraishi, and H.~Takayanagi,
Phys. Rev. B {\bf 65}, 081307 (2002);
M.~Ichimura, K.~Kusakabe, S.~Watanabe, and T.~Onogi, 
Phys. Rev. B {\bf 58}, 9595 (1998); 
H.~Ishii, T.~Nakayama, and J.-i.~Inoue, 
Phys. Rev. B {\bf 69}, 085325 (2004).

\bibitem{nishino-3d-flatband}
S.~Nishino, M.~Goda, and K.~Kusakabe, 
J. Phys. Soc. Jpn. {\bf 72}, 2015 (2003); 
S.~Nishino and M.~Goda, 
J. Phys. Soc. Jpn. {\bf 74}, 393 (2005).

\bibitem{flat-band-experiment-polymers}
R.~Arita, Y.~Suwa, K.~Kuroki, and H.~Aoki, 
Phys. Rev. Lett. {\bf 88}, 127202 (2002); 
Y.~Suwa, R.~Arita, K.~Kuroki, and H.~Aoki, 
Phys. Rev. B {\bf 68}, 174419 (2003);
H.~Aoki,
Applied Surface Science {\bf 237}, 2 (2004).

\bibitem{lin-grapheneribbon}
H.-H.~Lin, T.~Hikihara, H.-T.~Jeng, B.-L.~Huang, C.-Y.~Mou, and X.~Hu, 
Phys. Rev. B {\bf 79}, 035405 (2009).

\bibitem{flat-band-graphene}
M.~Garnica, D.~Stradi, S.~Barja, F.~Calleja, C.~Diaz, M.~Alcami,
N.~Martin, A.~L.~V.~de~Parga, F.~Martin, and R.~Miranda,
Nat. Phys. {\bf 9}, 368 (2013).

\bibitem{bloch2005}
I.~Bloch, 
Nat. Phys. {\bf 1}, 23 (2005).

\bibitem{bloch2008}
I.~Bloch, J.~Dalibard, and W.~Zwerger, 
Rev. Mod. Phys. {\bf 80}, 885 (2008).

\bibitem{chern2013}
G.-W.~Chern, C.-C.~Chien, and M.~Di Ventra,
arXiv:1307.6128. 

\bibitem{tsunetsugu}
M.~E.~Zhitomirsky and H.~Tsunetsugu,
Phys. Rev. B {\bf 70}, 100403(R) (2004);
{\bf 75}, 224416 (2007).

\bibitem{unser_EPJB}
O.~Derzhko and J.~Richter,
Phys. Rev. B {\bf 70}, 104415 (2004);
Eur. Phys. J. B {\bf 52}, 23 (2006).

\bibitem{shtengel}
M.~Maksymenko, R.~Moessner, and K.~Shtengel,
arXiv:1401.6172.

\bibitem{villain} 
J.~Villain, R.~Bidaux, J.~P.~Carton, and R.~Conte,
J. Phys. {\bf 41}, 1263 (1980).

\bibitem{shender} 
E.~F.~Shender,
Zh. Eksp. Teor. Fiz. {\bf 83}, 326 (1982)
[Sov. Phys. JETP \textbf{56}, 178 (1982)].

\bibitem{lin_indep}
H.-J.~Schmidt, J.~Richter, and R.~Moessner,
J. Phys. A. {\bf 39}, 10673 (2006).

\bibitem{azurite}
H.~Kikuchi, Y.~Fujii, M.~Chiba, S.~Mitsudo, T.~Idehara, T.~Tonegawa, K.~Okamoto, T.~Sakai, T.~Kuwai, and H.~Ohta,
Phys. Rev. Lett. {\bf 94}, 227201 (2005);
H.~Jeschke, I.~Opahle, H.~Kandpal, R.~Valenti, H.~Das, T.~Saha-Dasgupta, O.~Janson, H.~Rosner, A.~Br\"{u}hl,
B.~Wolf, M.~Lang, J.~Richter, S.~Hu, X.~Wang, R.~Peters, T.~Pruschke, and A.~Honecker,
Phys. Rev. Lett. {\bf 106}, 217201 (2011).

\bibitem{klein}
D.~J.~Klein, 
J. Chem. Phys. {\bf 61}, 786 (1974).

\bibitem{fulde}
P.~Fulde,
{\it {Electron Correlations in Molecules and Solids}}
(Springer-Verlag, Berlin, Heidelberg, 1993),
p.~77.

\bibitem{essler}
F.~H.~L.~Essler, H.~Frahm, F.~G\"{o}hmann, A.~Kl\"{u}mper, and V.~E.~Korepin,
{\it {The One-Dimensional Hubbard Model}}
(Cambridge University Press, Cambridge, UK, 2005), 
p.~38.

\bibitem{harris}
A.~B.~Harris and R.~V.~Lange,
Phys. Rev. {\bf 157}, 295 (1967).

\bibitem{diamondchains_ssrealizations}
X.~Mo,  K.~M.~S.~Etheredge,  S.-J.~Hwu, and Q.~Huang,
Inorg. Chem. {\bf 45}, 3478 (2006);
R.~A.~Mole, J.~A.~Stride,  P.~F.~Henry, M.~Hoelzel, A.~Senyshyn, A.~Alberola, C.~J.~G.~Garcia, P.~R.~Raithby, and P.~T.~Wood,
Inorg. Chem. {\bf 50}, 2246 (2011).

\bibitem{honeck} 
M.~Ishii, H.~Tanaka, M.~Hori, H.~Uekusa, Y.~Ohashi, K.~Tatani, Y.~Narumi, and K.~Kindo, 
J. Phys. Soc. Jpn. {\bf 69}, 340 (2000);
A.~Honecker and A.~L\"{a}uchli,
Phys. Rev. B {\bf 63}, 174407 (2001).

\bibitem{spinpack}
{\tt{http://www-e.uni-magdeburg.de/jschulen/spin/}}

\end{thebibliography}
\end{document}